\documentclass[a4paper]{article}

\usepackage[english]{babel}
\usepackage[square,sort,comma,numbers]{natbib}

\usepackage{siunitx}

\usepackage[letterpaper,top=2cm,bottom=2cm,left=3cm,right=3cm,marginparwidth=1.75cm]{geometry}

\usepackage{amsmath}
\usepackage{graphicx}
\usepackage[colorlinks=true, allcolors=blue]{hyperref}

\usepackage{authblk}

\title{Contrast Enhancement of Barely Visible Impact Damage using Speckle-Based Dark-Field Radiography}

\author[1,2]{Ronan Smith}
\author[1]{Gregor Bortsnar}

\affil[1]{Faculty of Engineering and Physical Science, University of Southampton, University Road, Southampton, United Kingdom}

\affil[2]{Adelaide Medical School, University of Adelaide, North Terrace, Adelaide, Australia}

\affil[3]{Department of Physics, University of Trieste, A. Valerio 2, Trieste, Italy}

\affil[4]{Elettra Sincrotrone, Basovizza, Trieste, Italy}

\author[3,4]{Vittorio Di Trapani}
\author[3,4]{Ginevra Lautizi}

\affil[5]{European Synchrotron Radiation Facility, 71 Avenue des Martyrs, Grenoble, France}

\affil[6]{Institut Laue-Langevin, 71 Avenue des Martyrs, Grenoble, France}

\author[5]{Ludovic Broche}
\author[5,6]{Lukas Helfen}

\author[1]{Daniel Bull}
\author[1]{Richard Boardman}
\author[1]{S. Mark Spearing}
\author[1]{Ian Sinclair}
\author[1]{Mark Mavrogordato}
\author[3,4]{Pierre Thibault}

\begin{document}
\maketitle

\begin{abstract}
Barely visible impact damage (BVID) can cause serious issue for composite structures, due to sub-surface damage seriously reducing the strength of the material without showing easily detectable surface signs. Dark-field imaging measures ultra-small angle scattering caused by microscopic features within samples. It is sensitive to damage in composite materials which would otherwise be invisible in conventional radiography. Here we demonstrate BVID detection with speckle-based dark-field imaging, a technique requiring only sandpaper (to create the speckle-pattern) in addition to a conventional X-ray imaging setup to extract the dark-field imaging. We demonstrate that the technique is capable of detecting both matrix cracking and delaminations by imaging materials susceptible to these failure mechanisms.

\end{abstract}










\section*{Introduction}
\label{sec:sample1}

Low velocity impacts to Carbon Fibre Reinforced Polymer (CFRP) laminated panels can lead to a significant loss in mechanical properties while leaving no indication of the damage on the surface of the material \cite{Richardson1996}. This barely visible impact damage (BVID) can be detected using a number of destructive and non-destructive techniques \cite{Wronkowicz-Katunin2019}, such as ultrasound \cite{Wronkowicz2018} X-rays \cite{Garcea2018X-rayComposites}. In particular, X-ray radiographs which measure the attenuation of the X-ray beam as it passes through a sample can be used to detect damage in various materials. However, as the cracks are often small, and in BVID the material can be displaced parallel to the path of the X-rays (so there is no net density change along their path), damage can be hard to detect. Contrast agents can be introduced into the damage to highlight it \cite{Spearing1992}, however, this requires the damage be connected to the surface \cite{Yu2016} and makes subsequent repair of any damage difficult. Computed tomography (CT), by yielding three-dimensional images of the internal structure of samples can overcome the limitations of radiographic images in detecting delaminations. However, for large, plate-like objects, artefacts can be produced when the long axis of the specimen aligns with the beam \cite{Yu2016}. Computed Laminography (CL) avoids this by rotating the sample along an axis angled towards the beam \cite{Helfen2007Synchrotron-radiationDevices}, and has been shown to be effective at evaluating cracking in CFRP plates \cite{Moffat2010}. Further contrast within CFRP samples in X-ray radiography and tomography can be obtained by taking advantage of propagation-based edge-enhancement \cite{Cloetens1997a} or using optical elements \cite{Shoukroun2022EdgeCFRP}.

Dark-Field X-ray imaging can also be used to enhance contrast for damage in CFRP \cite{Endrizzi2015Edge-illuminationStructures, Senck2018, Rus2020}. Dark-Field imaging measures the diffusion or scattering of the X-rays caused by microstructures within the sample \cite{Pfeiffer2008}. This also makes it sensitive to resin rich areas \cite{Glinz2021} and changes in porosity \cite{Shoukroun2023}. It has also been used to study glass-fiber based materials \cite{ozturk2023}. Dark-field X-ray imaging requires only standard X-ray imaging systems, with the addition of optical elements to pattern the beam. Previous demonstrations for imaging CFRP have involved the use of custom X-ray gratings, arranged into a Talbot-Lau interferometer  (see \cite{Senck2018, Rus2020, Glinz2021}) or an Edge-Illumination system (see \cite{Endrizzi2015Edge-illuminationStructures, Shoukroun2022EdgeCFRP, Shoukroun2023}). Speckle-based imaging aims to measure the same signals using readily available sandpaper as the optical element instead \cite{Zdora2018}, allowing for low cost addition to existing X-ray imaging systems \cite{Zdora2020a}. 

In Speckle-Based Imaging, a diffuser (typically sandpaper) is placed in the X-ray beam upstream of the sample stage, creating a random intensity pattern. When the sample is introduced, the dark-field image is detected by looking at how the pattern has been blurred by the sample. This demonstration uses the Unified Modulated Pattern Analysis (UMPA) algorithm \cite{Zdora2017, DeMarco2023, smith2022} to extract these signals, see Zdora \cite{Zdora2018} for an in-depth comparison of other speckle-based imaging techniques. The UMPA algorithm is a versatile algorithm, being compatible with synchrotron and laboratory X-ray sources \cite{Zdora2020, smith2023}, as well as patterns from structured optical elements \cite{Smith2024-zv}. Samples larger than the field of view can be raster scanned, with the data producing a single large image \cite{DeMarco2023}.

BVID is not caused by a single damage mechanism, there are several modes of failure in play, as discussed by Bouvet et.al. \cite{bouvet2012}. Two of the main mechanisms are delamination of the plies and the formation of cracks within a ply running parallel to the fibres, known as matrix cracking. These cracks have different orientations and geometries, and so interact differently with an X-ray beam, giving different dark-field signals. To our knowledge, no previous studies have investigated whether dark-field X-ray imaging is capable of detecting both of these damage mechanisms, or whether the contrast results from one damage type alone. Understanding this is critical for dark-field X-ray imaging to becoming a more widely adopted technique in the field of non-destructive testing of composite materials. 

\section*{Materials and Methodology Part I - Samples and Damage Verification through Computed Laminography}
\label{sec:meth-cfrp}

\begin{figure}
    \centering
    \includegraphics[width=0.7\textwidth]{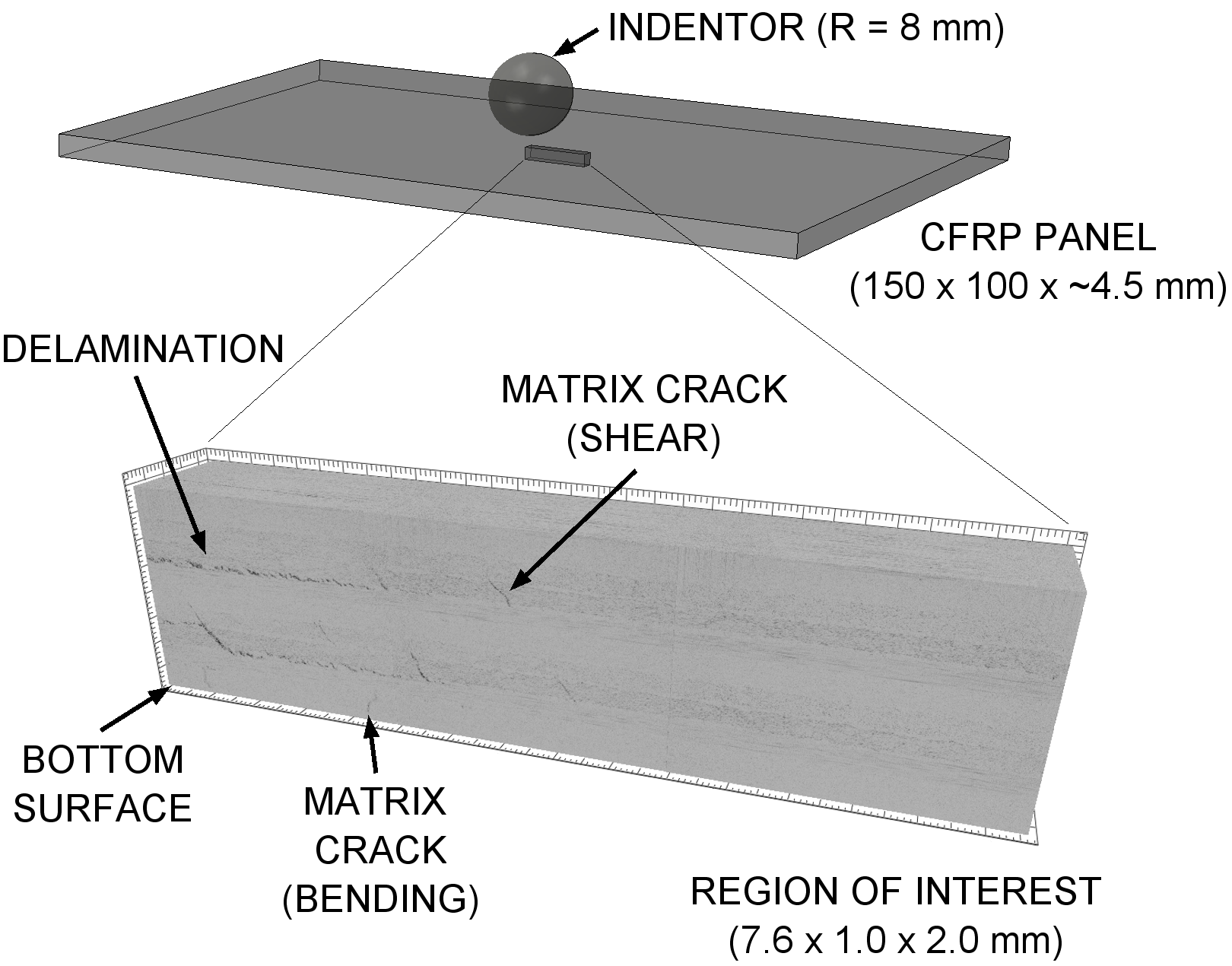}
    \caption{A diagram showing the approximate location of the imaging site relative to the indentation on the CFRP plate, with a reconstructed computed laminography volume showing the damage mechanisms seen in the delaminating material - note that some matrix cracking is present.}
    \label{fig:setup_g}
\end{figure}

As demonstrated by Bull et.al. \cite{Bull2014}, the addition of particles to the matrix of a composite material can be used to influence the damage mechanisms seen in BVID. Two materials were designed and manufactured using this approach, which showed either a delamination or matrix-cracking dominated failure mechanism when subjected to an out-of-plane quasi-static load. The samples were prepared by Solvay (Belgium), and for simplicity in this paper we refer to them as the delaminating material and matrix-cracking material respectively. 

The samples were prepared in accordance with the ASTM D7137M standards, using a 24 ply layup with a $[45/0/-45/90]_{3\textbf{s}}$ stacking sequence \cite{ASTM}. The thickness of the cured composite sheet was \SI{4.5}{\milli \metre} $\pm$ \SI{0.2}{\milli \metre}, coupons were waterjet cut and end-mill finished into \SI{150}{\milli \metre} $\times$ \SI{100}{\milli \metre} sections. Two different types of secondary phase thermoplastic particles were dispersed in the interlayers, forming 13\% of the pre-cure weight of the interlayer. The base-epoxy resin and intermediate modulus carbon fibre was the same between both materials. The exact composition of the thermoplastics added are proprietary and do not effect the findings of this paper. 

Performance of each material was tested in accordance with the EN 6034 and ASTM D7136M \cite{ASTM}. The delaminating material had  \(\sim \)80\% of the G\textsubscript{IIC} value and compression after impact (CAI) failure stress than the matrix cracking material.

To induce the required damage for imaging, the samples were laid over a \SI{125}{\milli \metre} $\times$ \SI{75}{\milli \metre} window, and subject to a through-thickness load, applied to the centre of the plates. A \SI{4}{\milli \metre} displacement of the indentor (radius \SI{8}{\milli \metre}) was applied. 

To confirm the presence and mechanisms of damage seen in the samples, synchrotron Radiation Computed Laminography (SRCL) was employed. In SRCL, the sample is rotated about an axis inclined to the X-ray beam, allowing for less artefacts when scanning plate-like samples than would be possible with computed tomography (CT) \cite{Helfen2007, Helfen2011}. The ID19 beamline of the European Synchrotron Radiation Facility (ESRF) was used, with an average photon energy of \SI{19}{\kilo \electronvolt}. Scans were taken using 2400 projection images with \SI{100}{\milli \second} exposure time. The pixel size was \SI{1.4}{\micro \metre} in the sample plane, with a total field of view of \SI{2.8}{\milli \meter} $\times$ \SI{2.8}{\milli \meter}. Four adjoining scans were performed on each material, giving reconstructed volumes of \SI{7.6}{\milli \meter} x \SI{2.0}{\milli \meter} x \SI{1.0}{\milli \meter} along the bottom surface of the specimen. These parameters were chosen due to their previous successful use \cite{Borstnar2015}. 

Figure \ref{fig:setup_g} shows the resulting volume for the delaminating material, with delamination between the plies clearly visible. A matrix crack within one of the plies can also be seen (as the damage mechanisms exhibited in the two materials manufactured is not mutually exclusive). The damage in these volumes was manually segmented. 

\section*{Materials and Methodology Part II - Dark-Field Imaging}
\label{sec:meth-df}

Dark-field radiography was performed at the ID19 beamline of the ESRF synchrontron. A diffuser, made by clamping together a stack 6 sheets of P180 grit silicon-carbide sandpaper, placed \SI{1.89}{\metre} upstream of the sample was used to create the speckle pattern. The sample was \SI{6.65}{\metre} from the \SI{80}{\micro \metre} Gadox scintillator, coupled to a PCO.edge SCMOS detector with a magnifying objective. This gave a pixel size of \SI{21.6}{\micro \metre} in the sample plane. A polychromatic beam with mean photon energy \SI{37.7}{\kilo \electronvolt} was produced by a wiggler set to a gap of \SI{100}{\milli \metre}, and filtered using \SI{10}{\milli \metre} of aluminium and \SI{0.14}{\milli \metre} of copper. These parameters lead to the size of the features in the speckle pattern produced by the sandpaper was on the order of several pixels on the detector. 6 sheets of sandpaper produced a more visible pattern than with a single sheet, while not attenuating the beam too heavily. 

The beam profile meant that the usable field of view on the detector was 1930 $\times$ 370 pixels, or \SI{41.6}{\milli \meter} $\times$ \SI{7.99}{\milli \meter}. The sample was stepped through the beam in a grid pattern, with 560 frames taken in total giving a field of view of 5688 $\times$ 6322 pixels, or \SI{122}{\milli \metre} $\times$ \SI{137}{\milli \metre}. These frames overlapped so that the area of the sample was covered in at least 10 frames, as the information from multiple images can be incorporated to increase the signal to noise ratio \cite{smith2022}. Fifty \SI{5}{\milli \second} exposures were taken at each sample position, and averaged to reduce noise. This lead to to a total exposure time of \SI{140}{\second} for the 560 images taken of the sample. A reference image of the sandpaper without the sample present is also required, which again was the average of 50 \SI{5}{\milli \second} exposures. Reference patterns were taken after every move of the sample in case stability was an issue, however this was not the case.

The dark-field images were extracted using the UMPA algorithm \cite{Zdora2018, DeMarco2023} This algorithm works by comparing a reference image of just the speckle pattern (sandpaper alone in the beam) to the sample frames (sandpaper and sample in the beam). At every pixel in the image, the transformation of a window (with size 7 $\times$ 7 pixels in our case) between the sample and reference frames is calculated. The dark-field image is found by looking at how the speckle pattern has been blurred, and the transmission by the loss of intensity within the window.  

\begin{figure}
    \centering
    \includegraphics[width=0.95\textwidth]{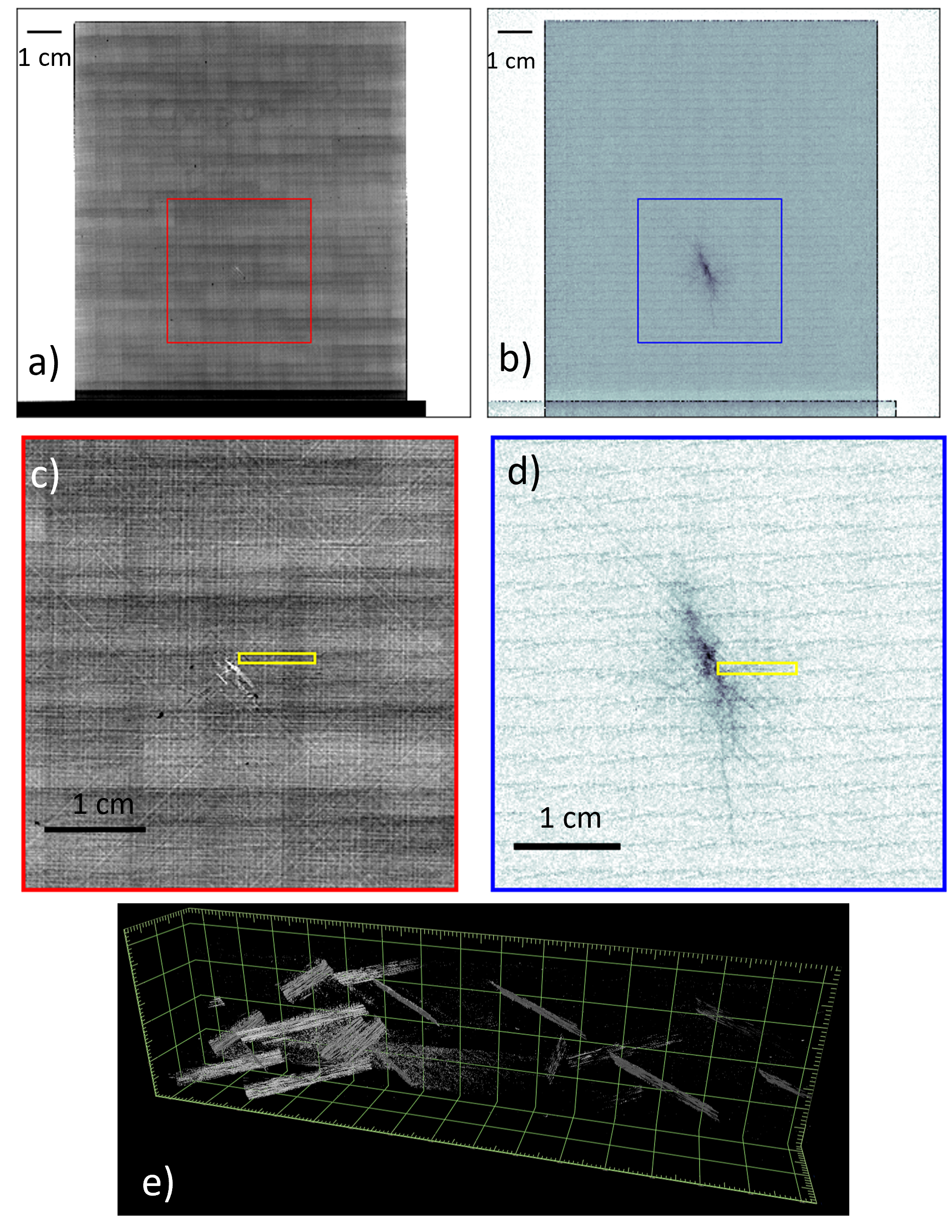}
    \caption{Transmission (a) and dark-field (b) radiographs of the whole sample of the matrix-cracking material. The red and blue coloured rectangles indicate the regions of interest shown in (c) and (d) with the colour-maps changed to enhance the damage visibility. The yellow rectangles on the region of interest images show the approximate locations of the computed laminography scans. The segmented damage from this computed laminography volume (e) shows mostly matrix cracking.}
    \label{fig:result_matb}
\end{figure}

\section*{Results and Discussion}
\label{sec:results}

The results of the dark-field and CL imaging for the matrix-cracking material are shown in figure \ref{fig:result_matb}. Dark-field imaging produces a transmission as well as a dark-field image. This transmission image measures transmission in the sample, and is akin to what the imaging system would have produced if it was operating as a normal X-ray radiography setup. In the overview scan (Fig \ref{fig:result_matb}a and Fig \ref{fig:result_matb}b), the contrast in the transmission image has been set to differentiate the CFRP from the aluminium holder and the air. The faint horizontal and vertical lines in the transmission and dark-field images are due to fluctuations of the X-ray beam intensity in the individual frames that were taken to create the images shown. The position of the beam was also fluctuating slightly on the detector, making any correction for this impractically challenging. Regions of interest around the damage are shown in figures \ref{fig:result_matb} c) and \ref{fig:result_matb} d), in these the contrast has been further increased. This makes the damage in the transmission image just visible, and also improves the damage visibility in the dark-field image. The increased contrast in the dark-field images makes the crack running from the bottom of the impact site much clearer. 

\begin{figure}
    \centering
    \includegraphics[width=0.95\textwidth]{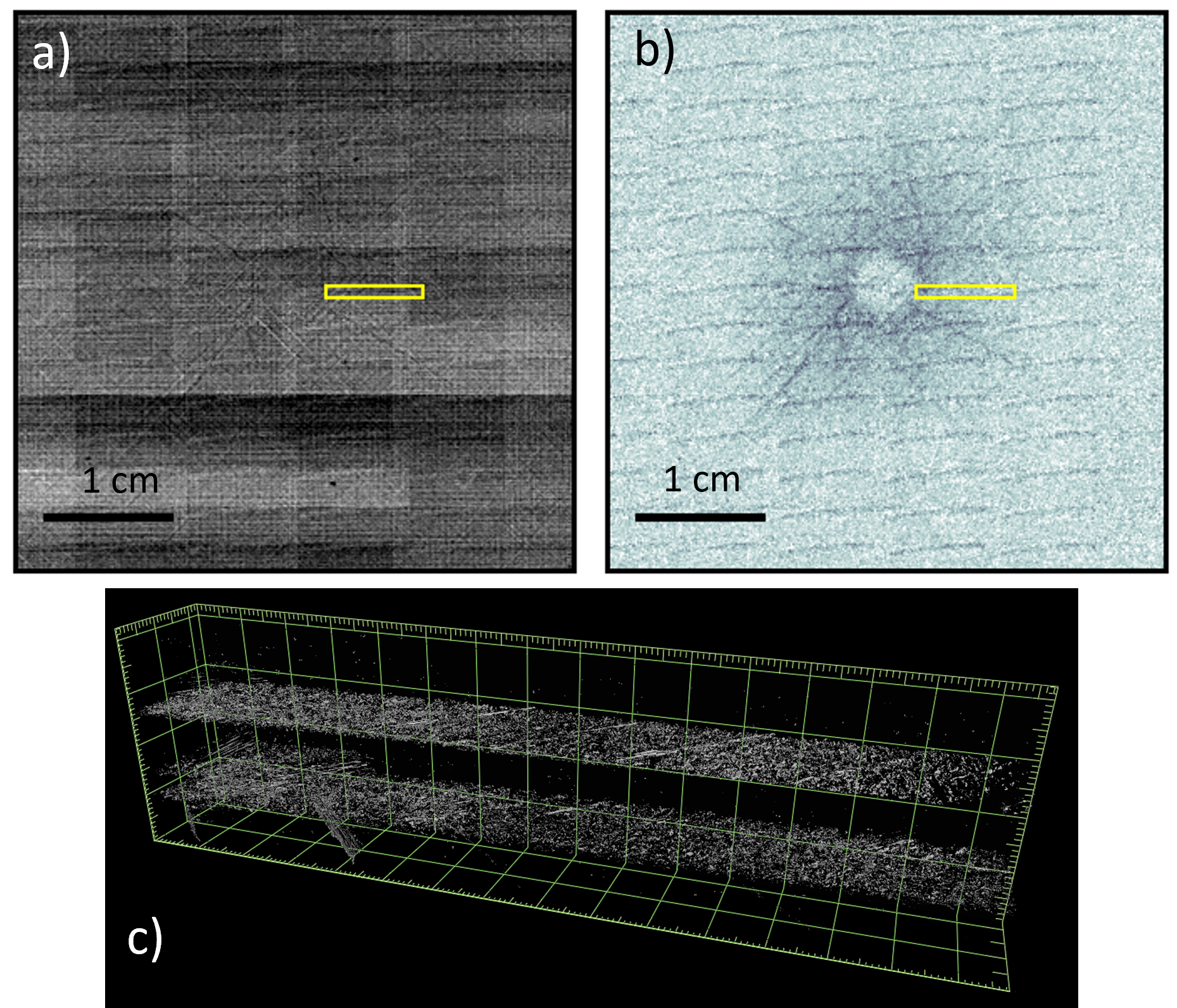}
    \caption{Transmission (a) and dark-field (b) radiographs of the delaminating material, these are the same size as the region-of-interest images shown above in figure \ref{fig:result_matb} for the matrix-cracking material. The yellow rectangles on the region of interest images show the approximate locations of the computed laminography scans. The segmented damage from this volume (c) shows mostly delaminations.}
    \label{fig:result_mata}
\end{figure}

For the delaminating material (figs \ref{fig:result_mata} a) and \ref{fig:result_mata} b), we could not find a contrast level which made the damage visible in the transmission image. However, it is clear in dark-field. The difference between the dark-field images produced by the two materials is striking. For the delaminating material, the region directly under the impact site appears undamaged, with a halo of damage surrounding this region. A few cracks appear to be emanating from this halo. The matrix-cracking material has developed a large crack centred under the indentation site, again with several cracks propagating away from it. 

The yellow boxes shown in the figures indicate the approximate regions in which the segmented CL volumes show in in figs \ref{fig:result_matb} e) and \ref{fig:result_mata} c) were taken. The segmented volume shows that following the out-of-plane loading, the matrix cracking material (fig \ref{fig:result_matb} e) was more effective at suppressing delaminations, which have been shown to propagate under Mode-II dominated loading conditions and is consistent with the mateials higher Giic. Figure \ref{fig:result_mata} c) shows that segmented damage in the delamination material was mainly localised along two planes between the plies of the laminate, corresponding to delaminations.
One matrix crack is visible in the left hand side of the lower delamination layer in the volume. This may correspond to one of the cracks visible in the dark-field image (fig \ref{fig:result_mata} d). The matrix-cracking material has many more matrix cracks (fig \ref{fig:result_matb} e), caused by an increased toughness supressing delamination. Spatially, these are at a much higher density on the left-hand side of the volume, closer to the impact site, with a few smaller regions of damage further from the site. Due to the very strong dark-field signal given off at the very centre of the damaged region, the displayed contrast level used in the dark-field image for materials B (fig \ref{fig:result_matb} d) makes seeing the effect of the few smaller matrix cracks at the right hand side of the segmented volume challenging. 
This shows that speckle-based dark-field imaging (as well as other dark-field techniques) are capable of detecting damage caused by both matrix cracking and delamination in BVID.

Our use of speckle-based imaging shows that dark-field visualisation of BVID is possible without the use of customised optical elements employed in other dark-field imaging techniques. This technique is easily implementable at synchrotron imaging facilities worldwide, requiring no complex alignment procedures and only a sheet of sandpaper. Additionally, speckle-based dark-field imaging using the UMPA algorithm has been demonstrated compatible with laboratory sources as well \cite{Zdora2020, smith2023}, potentially allowing for this technique to be integrated into lab-based non-destructive testing systems in future. The use of a polychromatic beam in this experiment highlights that a monochromatic beam is not required, however it is worth noting that a typical lab source would give a much broader spectrum of energies than the beam we used in the present work. The use of the UMPA technique of speckle-based imaging and its capabilities to stitch together frames with the sample in multiple positions to create a single image means the size of sample is limited only by the range of travel on the motors used to position the sample. As with CT \citep{GalvezHernandez2023}, an optimisation of parameters could yield better results for imaging CFRP samples.  Work towards dark-field imaging without optical elements at synchrotron sources is ongoing \cite{Leatham, Ahlers:24}. Recently a single-shot technique exploiting monochromater energy harmonics alongside an energy-discriminating detector has been demonstrated \cite{Ahlers:25}, potentially allowing for optics-free dark-field X-ray imaging to be used for BVID detection in future. 

\section*{Conclusion}

We have successfully shown the dark-field radiography can effectively visualise barely visible impact damage. By imaging samples susceptible to matrix cracking and delamination, and verifying that these are the main damage mechanisms in each sample, we have show that dark-field radiography is sensitive to both damage mechanisms. 

This was the first demonstration of visualising barely visible impact damage in dark-field using a speckle-based imaging techniques, showing that it is possible without the need for customised X-ray optical elements used in previous dark-field imaging demonstrations.

\label{sec:conclusion}

\section*{Acknowledgements}

We acknowledge the European Synchrotron Radiation Facility (ESRF) for provision of synchrotron radiation facilities and we would like to thank Ludovic Brochs and Lukas Helfen for assistance and support in using beamline ID19. We thank Solvay for providing the samples. 

We acknowledge funding from the European Research Councils Horizon 2020 Consolidator Grant project 'Scattering-based X-ray Imaging and Tomography'.

\section*{Data Availability}

A copy of the data can be found at \cite{CFRP_data}, and a copy of the UMPA code used to analyse it can be found at \cite{fabio_de_marco_2022_6984740}. 

 \bibliographystyle{elsarticle-num} 
 \bibliography{references, sample}

\end{document}